\documentclass[a4paper]{jpconf}
\usepackage{amsmath,amssymb}
\usepackage{hyperref}
\usepackage{graphicx}
\pdfoutput=1

\begin{document}
\title{Characterisation of diffusion-driven self-organisation of rodlike particles by means of entropy of~generalised two-dimensional words}

\author{Mikhail~V~Ulyanov$^{1,2}$, Yuri~G~Smetanin$^{3,4}$, Mikhail~M~Shulga$^5$, Andrei~V~Eserkepov$^6$, Yuri~Yu~Tarasevich$^6$}
\ead{tarasevich@asu.edu.ru}
\address{$^{1}$V\,A\,Trapeznikov Institute of Control Sciences of RAS, 
Moscow, Russia}
\address{$^{2}$Computational Mathematics and Cybernetics, M\,V\,Lomonosov Moscow State University, Moscow, Russia}
\address{$^{3}$Moscow Institute of Physics and Technology, Moscow, Russia}
\address{$^{4}$Federal Research Center ``Informatics and Control'' of RAS, 
Moscow, Russia}
\address{$^{5}$Higher School of Economics, Moscow, Russia}
\address{$^{6}$Astrakhan State University, Astrakhan, 414056, Russia}

\begin{abstract}
The experiments conducted by various scientific groups indicate that, in dense two-dimensional systems of elongated particles subjected to vibration, the pattern formation is possible. Computer simulations have evidenced that the random walk of rectangular particles in a discrete two-dimensional space can lead to their self-organisation.
We propose a technique for calculating the entropy characteristics of a two-dimensional system in a discrete two-dimensional space consisting of rectangular particles of two mutually perpendicular orientations, and a change in these characteristics for a random walk of particles is investigated.
\end{abstract}

\section{Introduction}\label{sec:intro}

During recent years, the attention of researchers has been drawn to the processes of self-organisation in dense two-dimensional (2D) granular systems~\cite{Aranson2006RMP}. In particular, many experimental works have been devoted to the pattern formation in thin layers composed of elongated granules subjected to vibrations~\cite{Boerzsoenyi2013SM,Muller2015PRE,Gonzalez2017SM}.

To model such the systems, a lattice approach is convenient to utilise, viz., a discrete space (a square lattice) is used; the rod with the aspect ratio $k$ is represented as a linear $k$-mer (rectangle of size $1 \times k$ lattice units). In this case, the rod can have only two mutually perpendicular spatial orientations, and the coordinates of its angles can be only integer.

The diffusion of $k$-mers on a square lattice has been studied using the kinetic Monte Carlo (MC) simulation~\cite{Lebovka2017PRE,Tarasevich2017JSM,Tarasevich2018JPhCSpacking}. Initial states with desired packing fraction of $k$-mers were produced using the random sequential adsorption (RSA)~\cite{Evans1993RMP}. Since the $k$-mers were considered as hard-core (completely rigid) particles, their reorganisation should necessarily be entropy-driven. Hard-core interaction between the rods means that energy of interparticle interaction, $U$, is defined as
\begin{equation}\label{eq:HCI}
  U =
  \begin{cases}
    0, & \mbox{if } r>d, \\
    \infty, & \mbox{if } r \leq d,
  \end{cases}
\end{equation}
where $r$ is the distance between two particles and $d$ is the  lattice constant.

The Helmholtz free energy, $F$, is defined as
 $ F = U - TS,$
where $U$ is the internal energy, $T$ is the absolute temperature  of the surroundings, and $S$ is the entropy of the system. At equilibrium, the Helmholtz free energy of a system at constant temperature and volume is minimal. Due to hard-core interaction between the rods~\eref{eq:HCI}, the internal energy, $U$, equals to zero. When the temperature is constant, any changes of the free energy, $F$, can be induced only by changes of the entropy, i.e.,
 $ \Delta F =  - T \Delta S.$
Pattern formation or phase transitions in such the system are said to be entropy-driven. In this case, an increase in macroscopic order is driven by an increase of microscopic disorder, i.e., a particle has more free volume to move in the final ordered state than it had in the initial disordered state~\cite{Frenkel1993}.

The main attention has been paid to isotropic systems, i.e., to systems with an equal number of vertical and horizontal particles~\cite{Lebovka2017PRE,Tarasevich2017JSM}. In dense systems, only translational diffusion of particles is possible, while rotational diffusion is completely suppressed. The system is isolated, i.e., neither matter nor energy can come in and go away. The influence of the aspect ratio of the particles, their concentration, the size of the system, and the type of boundary conditions on the formation of structures and the change in physical properties have been studied~\cite{Lebovka2017PRE,Tarasevich2017JSM}.

There are two possibilities of random walk~\cite{Mitescu1983,Selinger1990PRA}. In the first case, the particles are ``blind'', i.e., a particle chooses one of the four possible directions and tries to move in this direction. If the attempt is unsuccessful, the particle does not try to find other direction available for its movement. Such the behavior obeys detailed balance condition~\cite{Patra2018PRE}. In the second case, a particle is ``myopic'', i.e., it consistently tries in a random order all four possible direction until the first successful attempt to move or until all possibilities are exhausted~\cite{Lebovka2017PRE}. This behavior violates detailed balance condition~\cite{Patra2018PRE}. Such the particle can be called ``intellectual'' because it chooses with equal probability one of the possible directions to move if any. ``Myopic'' or ``intellectual'' particles may be related to active colloids.

Let us suppose, that there are $N$ particles in the system under consideration. An attempted displacement of the total number of particles in the system under consideration, $N$, is called one MC step or MC time unit. There are two possibilities to random choice of the next particle~\cite{LandauBinder}. The first possibility is to perform a random permutation of all $N$ particles and then consistently to go through these randomly reordered particles. In this case, each particle is chosen one and only one time at each MC step. The second possibility is a selection with return. In this case,  at the current MC step, some particles may be chosen several times but some particles may not be chosen at all. Both algorithms lead to the same final states but intermediate dynamics of the systems may differ. The latter algorithm was used in~\cite{Lebovka2017PRE,Tarasevich2017JSM}.

\begin{figure}
\begin{minipage}[b]{0.6\textwidth}
  \centering
($a$)\includegraphics[width=0.4\textwidth]{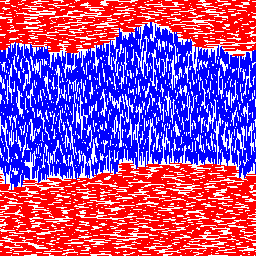}
\hfill
($b$)\includegraphics[width=0.4\textwidth]{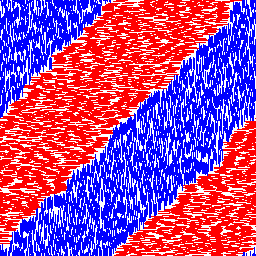}
\end{minipage}\hfill
\begin{minipage}[b]{0.3\textwidth}
\caption{Examples of patterns; $L=256$, PBCs, $t_\text{MC} = 10^7$, $k=12$.  The initial states were isotropic and jammed. ($a$) ``blind'' particles; ($b$) ``myopic'' particles}
\label{fig:patterns}
\end{minipage}
\end{figure}

Simulations evidenced that random walk of ``myopic'' particles in a dense system subjected to periodic boundary conditions (PBCs) produces pattern formation in a form of stripe domains~\cite{Lebovka2017PRE,Tarasevich2017JSM}, when the aspect ration of the particles is greater than 5 and number of MC steps is of order $10^6$ for lattice size $L=256$. The relaxation time  significantly depends on lattice size~\cite{Lebovka2017PRE,Tarasevich2017JSM}. Pattern formation in a system of ``blind'' particles occurs for longer particles~\cite{Patra2018PRE}. \Fref{fig:patterns} demonstrates examples of patterns. The pattern formation resembles the spinodal decomposition if $k$-mers oriented along the $x$ and $y$ directions ($k_x$-mers and $k_y$-mers, respectively) are treated as two ``phases''. If this similarity reflects the intrinsic nature of the process, then the pattern formation presumably  have to be described by the Cahn--Hilliard equation.

The configurational entropy has been applied to analyze entropy-driven phase transition in a system of long rods on a square lattice~\cite{Linares2008JSM}. Recently, the orientational phase transitions that occur in the deposition of $k$-mers  were characterised by information theory techniques~\cite{Vogel2017PRE}. Namely, the data recognizer ``word length zipper'' (wlzip) was created and utilised to find repeated meaningful information in any sequence of data.

Despite the fact that the changes in various characteristics with time has been studied and reported~\cite{Lebovka2017PRE, Tarasevich2017JSM}, the key value---entropy---has not been investigated yet. Since the extent of the disorder of the system is characterised by its entropy, the study of this particular quantity is expected to be most important for understanding the details of diffusion-driven self-organisation. In this paper, we propose to apply the entropy approach to the study of the process of self-organisation, i.e. the study of the change in the entropy of such a system over time.

The rest of the paper is constructed as follows. We describe mathematical background of our research in~\sref{sec:methods}. \Sref{sec:results} presents our main findings in dynamics of entropy. \Sref{sec:concl} summarizes the principal results.

\section{Methods}\label{sec:methods}
In our consideration, we only deal with very dense systems and square regions. The initial concentration of particles corresponds to the jamming, i.e. the state when no one additional particle can be added to the system because all presented voids are too small or their have unsuitable shapes. In such the system, only longitudinal displacement of a particle is obviously  allowed whilst any transversal displacement is undoubtedly forbidden.

In the present study, we utilize three main quantities, i.e., the normalised degree of freedom (DOF) per particle, $f$, the order parameter, $s$, and the information entropy, $H$.

We define the DOF as number of possible unit movements of a particle from its current location. The normalised DOF is defined as current value of DOF divided by its value at $t_\text{MC} = 0$.

The order parameter is defined as
\begin{equation*}\label{eq:s}
s_l = \frac{\left|N_y - N_x\right|}{N},
\end{equation*}
where $N_x$ and $N_y$  are the numbers of $k_x$-mers and $k_y$-mers, respectively, and $N=N_y + N_x$ is the total number of $k$-mers. This order parameter, $s_l$, was calculated in a sliding window of $l \times l$ sites and averaged over the entirely set of windows, i.e., over $L^2$ windows except empty windows. We utilize $l=2^n$, where $n = 1,2,\dots, 7$. In all cases, we considered only isotropic systems, i.e., $k_x$-mers and $k_y$-mers were equiprobable in their  deposition, hence, $s_l=0$ when $l=L$.

The entropy of finite one-dimensional words has been successfully used to study plant DNA~\cite{Smetanin2016MBB}, however, for 2D systems, a transition to 2D words and a corresponding modification of the method for calculating entropy values are necessary. Moreover, to study the dynamics of 2D words that encode the diffusion of particles on a square lattice with the use of entropy characteristics, it is necessary to solve two additional problems, viz.,
\begin{itemize}
  \item determination of the base of the logarithm for calculating the value of the entropy function,
  \item determination of the linear size, $l$, of a 2D window $l\times l$ sliding over the investigated 2D lattice $L \times L$ with a shift by one cell.
\end{itemize}

To calculate the entropy of the system at a particular MC step, we utilize the technique proposed  for one-dimensional words~\cite{Smetanin2016MBB}. Based on~\cite{Smetanin2016MBB}, we define the entropy of the 2D system under study at some MC step considering the original square lattice $L \times L$ as a 2D word, $w$, of size $L \times L$ in the three-character alphabet $\Sigma = \{A, B, C \}$, where $A$ encodes the empty elements of the lattice, a sequence of $k$ symbols $B$ corresponds to a horizontal $k$-mer, and a sequence of $k$ symbols $C$ corresponds to the vertical $k$-mer. The technique involves fixing the length, $l$, of a 2D window that sweeps along the lattice with a shift by one cell. In each position of the window, the observed 2D word is fixed and the frequency occurrence of the observed configurations over the entire lattice is calculated. The obtained frequencies are the basis for calculating the value of the entropy function of the 2D word under consideration using the formula for the entropy of discrete distributions
\begin{equation*}\label{eq:H}
H\left( w(t_\text{MC}),l,\alpha \right)=-\sum_{i=1}^{M}{\left( \frac{{\eta_i}}{m} \right)}{\log_\alpha}\left( \frac{{\eta_i}}{m} \right),
\end{equation*}
where $w$ is the 2D window under consideration, i.e., the lattice of size $L\times L$, $m$ is the total number of positions of the window $l\times l$ in the lattice (when the PBCs are applied, $m=L^2$), $\alpha$ is the base of logarithm, $M$ is the total number of different observed configurations, ${\eta_i}$ is the value of counter for $i$-th configuration, and $\sum_{i=1}^M \eta_i = m.$ Thus, the value of entropy function is determined, obviously, by the investigated 2D word, $w$, and by two parameters, viz., the base of the logarithm, $\alpha$, and the size of the window, $l$.

We denote MC steps as $t_\text{MC}$, where $t_\text{MC}=0,1,\ldots,10^7$. Then the investigated word $w$ corresponds to a certain MC step $t_\text{MC}$, i.e.,  $w = w\left(t_\text{MC} \right)$, for which the value of the entropy calculated as $H\left(w \left(t_\text{MC}\right), l, \alpha \right)$.

The main problem of direct application of the entropy calculation method for finite one-dimensional words~\cite{Smetanin2016MBB} to the 2D case is the over-exponential growth of the total number of unique configurations. In a 2D window of length $l$, the number of possible configurations for the three-character alphabet, in which the system under consideration is coded, is ${3^l}^2$. For example, in the window of size $ 5 \times 5 $, the total number of configurations is of the order of $ 8.5 \times 10^{11}$, although the total number of window positions in the lattice is only $256^2 = 65536$.

Our calculations evidenced that, starting from a window of length 5, the number of configurations observed in the lattice almost coincides with the number of window positions in the lattice, thereby such windows are generally not sensitive to changes in the lattice associated with the self-organisation process, and smaller windows show a weak entropy decrease as MC steps decrease, since they are sensitive to local lattice features. Thus, the problem arises of developing a modified method for calculating the entropy of 2D words that is sensitive to the process of self-organisation. It is necessary to move from local sensitivity to sensitivity with respect to global structural changes (i.e., over the entire lattice), to which the self-organisation of the particles actually leads. Such a transition is expected to be implemented by introducing generalised words of a three-character alphabet on the basis of equivalence classes.
We propose to consider equivalence classes of 2D words according to the number of alphabet symbols observed in the window.

This approach allows us to move from local features related to the arrangement of symbols in the window to generalised characteristics associated with the number of alphabets observed in the character window.

For the 2D word $v$ observed in a window of size $l \times l$, we introduce the function $N(v, u)$ whose value is the number of symbols $u$ in the window $v$. For the three-character alphabet under consideration $\Sigma = \{ A,B,C \}$, the equality
$
N(v,A) + N(v,B) + N(v,C) = |v|
$
is obviously valid.

Let us introduce an equivalence relation on 2D words of size $l \times l$, viz., two 2D words are equivalent if they have coincident quantities for all three symbols of the alphabet.
We denote as $W_l^{(a, b, c)}$ the equivalence class of two-dimensional words of size $l\times l$ with $a$ symbols $A$, $b$ symbols $B$, and $c$ symbols $C$
$
W_{l}^{(a, b, c)}=\left\{ v |N\left( v, A \right)= a, N\left( v, B \right) = b, N\left( v, C \right)= c \right\}.$
Thus, for example, all two-dimensional words of size $5\times 5$ that contain 2 symbols $A$, 12  symbols $B$, and 11 symbols $C$ fall into the equivalence class $W_5^{( 2,12,11) }$.
The number of such equivalence classes exactly coincides with the number of power representations of the window $l \times l$ as a sum of three nonnegative numbers. This number of representations is easily computed, since we need to put two delimiters in $l^2 + 2$ positions. Then the sum of the number of positions without delimiters is $l^2$, and the sum of the number of positions between the delimiters gives the desired representation. Consequently, the number of representations and the equivalent number of equivalence classes, $CE(l)$, is equal to
$$
CE(l) = \binom{l^2+2}{2} = \frac{1}{2} \left( l^4 +3 l^2 + 2 \right).
$$
For $ l = 5 $, there are only 351 generalised configurations instead of $8.5 \times 10^{11}$ unique configurations in the window, and, for $l = 6$, only 703 generalised configurations.
Thus, we modify the method for calculating the entropy of words, moving from calculating the frequency of occurrence of unique configurations to the frequency occurrence of elements of the factor of the set over the introduced equivalence classes.
Thus,
$$
\tilde H\left( w\left(t_\text{MC}\right),l,\alpha \right)=-\sum_{r=1}^{CE(l)}{\left( \frac{{\mu^{( a, b, c )}}}{m} \right)}{\log_\alpha}\left( \frac{\mu^{(a,b,c)}}{m} \right),
$$
where the value of the counter $\mu^{(a, b, c )}$ corresponds to the number of words in the lattice from the corresponding class $W_l^{(a, b, c)}$, the summation goes over all equivalence classes and the sum of the counters is equal to $\sum_{r=1}^{CE(l)}\mu^{( a, b, c)}=m.$

The base of the logarithm does not change the qualitative picture of temporal dynamics of entropy of generalised 2D words. Nevertheless, since self-information is determined by the logarithm to base 2, and the entropy is the probability-weighted self-information, we choice the binary logarithm. Thus, the entropy function $H\left (w \left(t_\text{MC}\right), l, \alpha \right)$ will be calculated using binary logarithm. Utilisation of the relative value of entropy, $H^\ast$, i.e., the ratio of the current value of the entropy, $H \left( w \left( t_\text{MC} \right), l, \alpha \right)$, to its initial value, $ H \left( w(0), l, \alpha \right) $, allows to omit the base.

\section{Results}\label{sec:results}

\begin{figure}
\begin{minipage}[b]{0.45\textwidth}
\centering
\includegraphics[width=\linewidth]{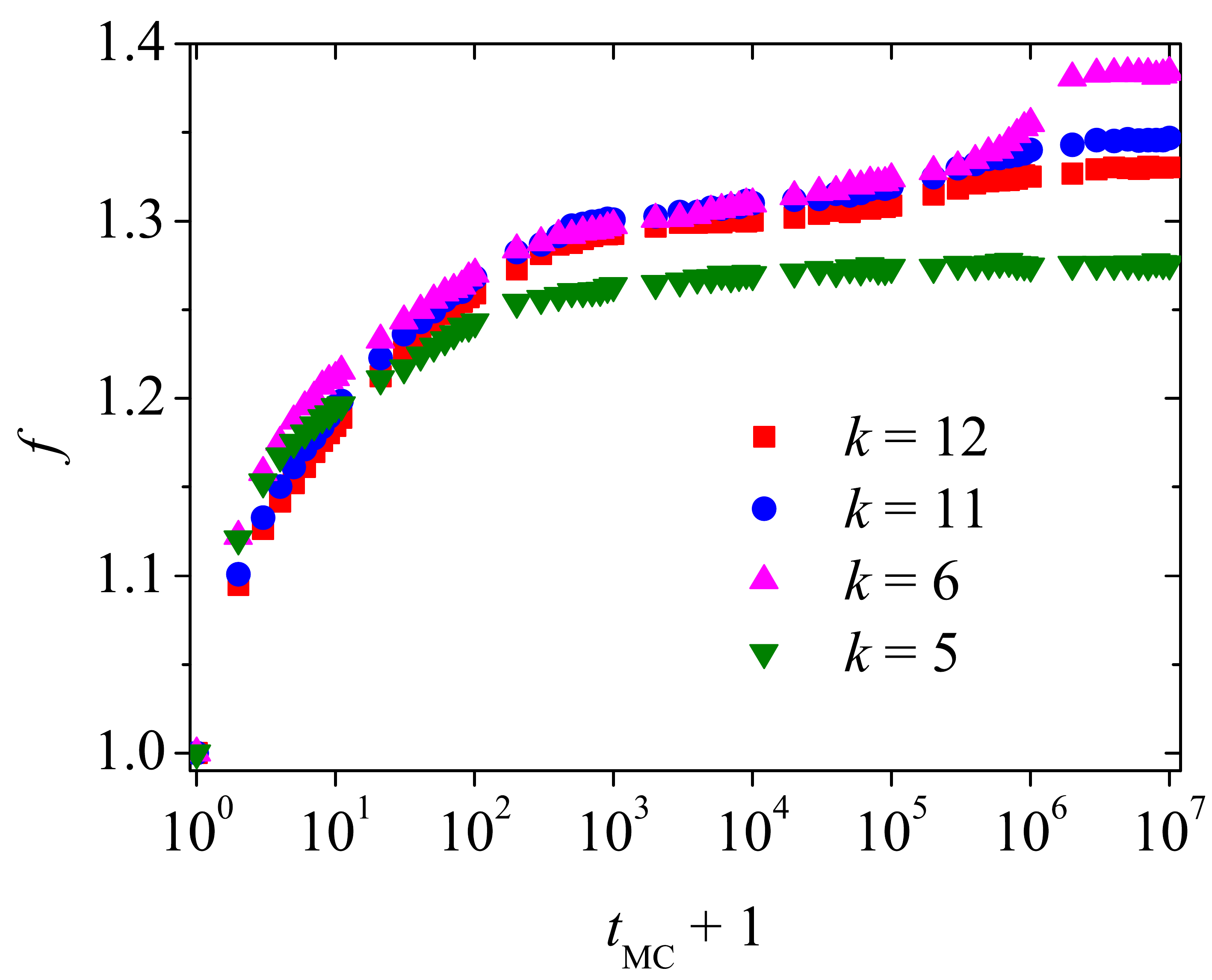}
\caption{Examples of normalised DOF, $f$, vs MC steps, $t_\text{MC}$, for different values of $k$ ($k=5,6,11,12$). The results were averaged over 100 independent runs.}
\label{fig:freedom}
\end{minipage}\hfill
\begin{minipage}[b]{0.45\textwidth}
\centering
\includegraphics[width=\linewidth]{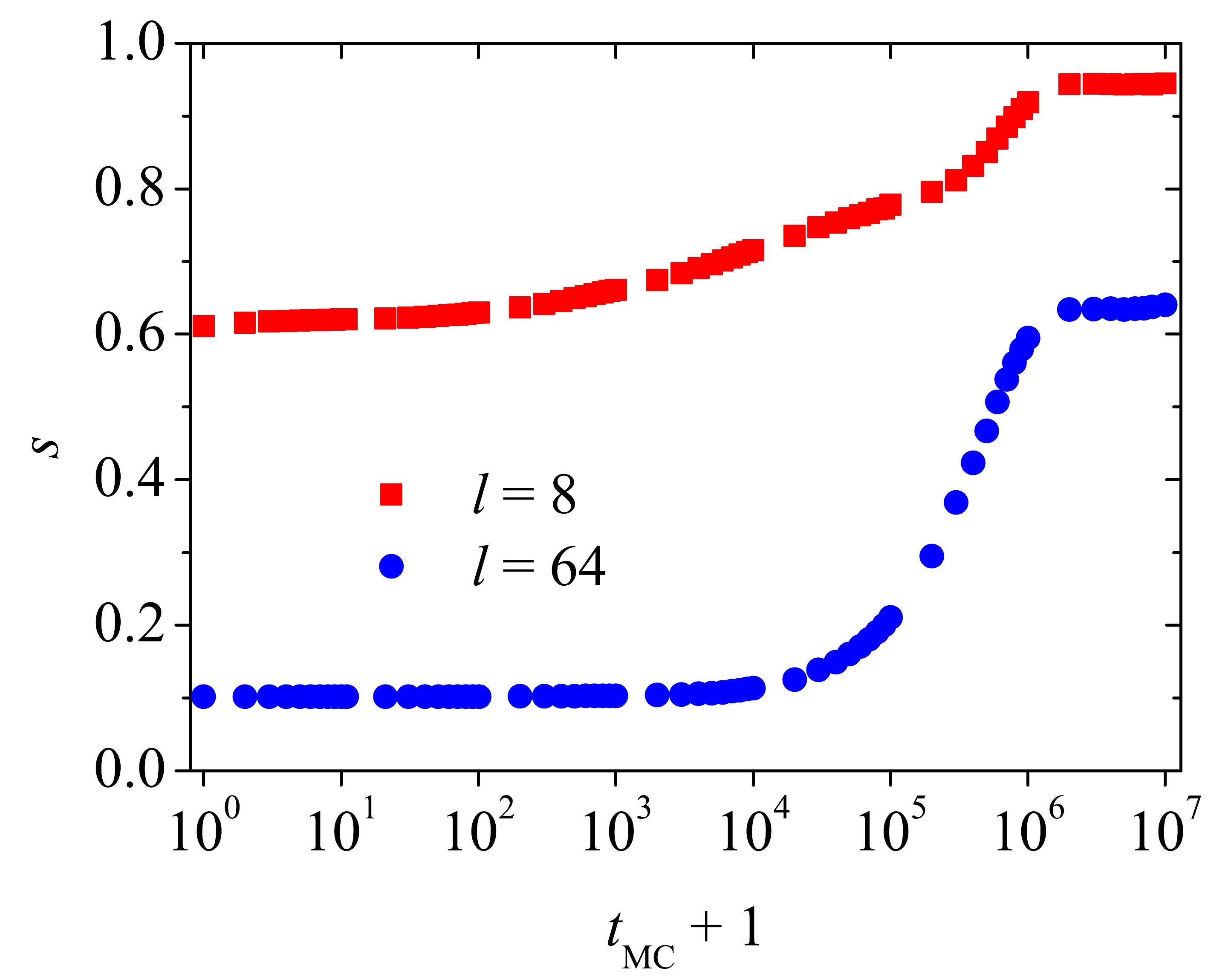}
\caption{Examples of local order parameter, $s_l$,  vs MC steps, $t_\text{MC}$, for different values of window size, $l = 8, 64$, and fixed value of $k$, $k=8$. The results were averaged over 100 independent runs.}
\label{fig:k8anisotropy}
\end{minipage}\hfill
\end{figure}

\Fref{fig:freedom} demonstrates how the normalised DOF per particle, $f$, varies with MC steps, $t_\text{MC}$. The curves evidenced that initial nonequilibrium state tends to equilibrate by means of random walk. In the final state, a particle has more free volume to move in its vicinity than it had in the jammed state. More significant growth of the DOF from 0 to $10^2$ MC steps corresponds to fluidisation~\cite{Tarasevich2017JSM}.

\Fref{fig:k8anisotropy} demonstrates how the local order parameter, $s_l$, varies with MC steps, $t_\text{MC}$. The curve corresponding to the window size $l=64$, i.e., $l= L/4$, evidenced that between $t_\text{MC} \sim 10^5$ and $t_\text{MC} \sim 10^6$, large clusters are forming. These clusters transform into stripe domains similar to presented in \fref{fig:patterns}($b$) when $t_\text{MC} \gtrsim 10^6$.

\begin{figure}[htpb]
\begin{minipage}[c]{0.45\textwidth}
\centering
\includegraphics[width=\linewidth]{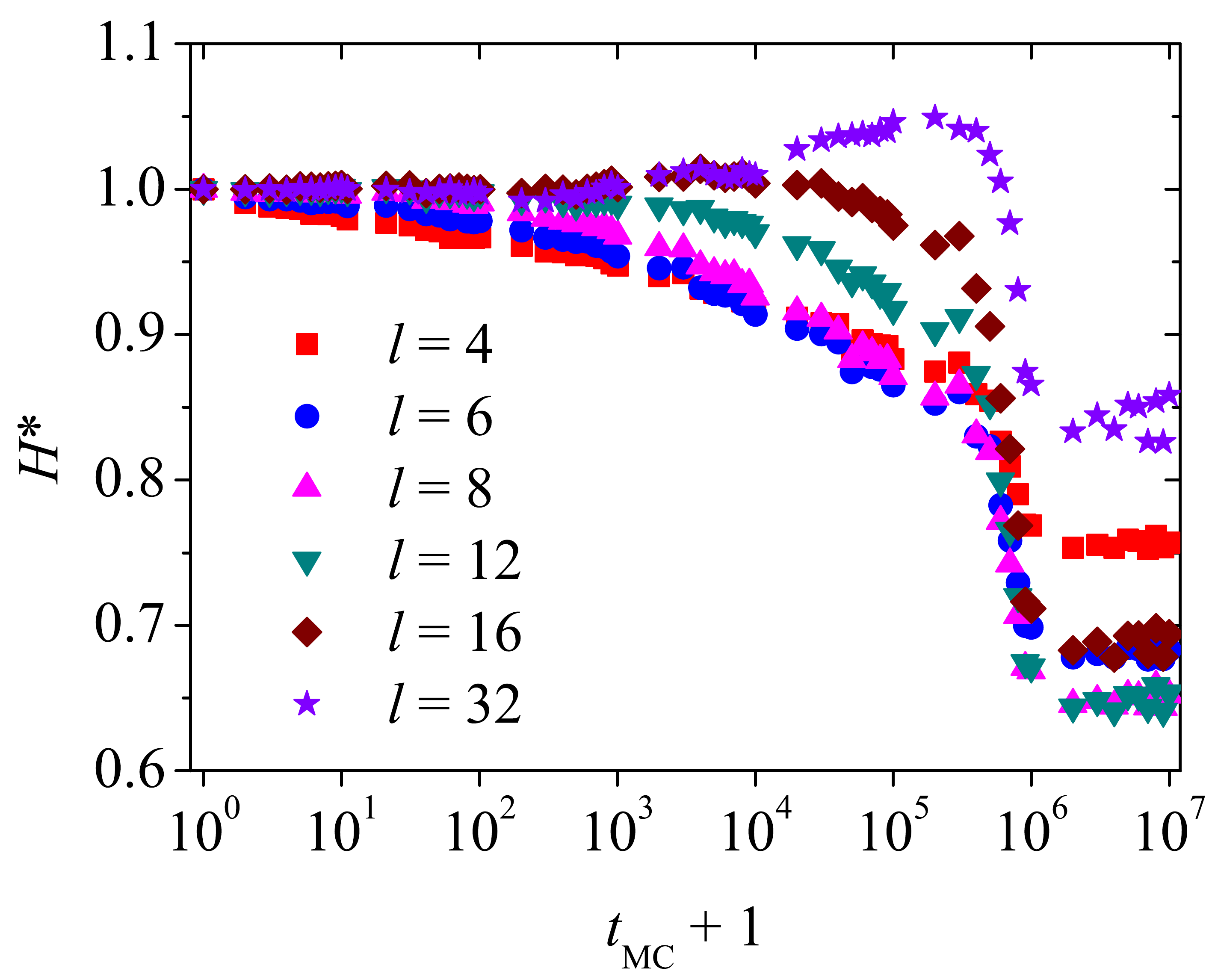}
\end{minipage}\hfill
\begin{minipage}[c]{0.45\textwidth}
\caption{Examples of normalised entropy, $H^\ast$,  vs MC steps, $t_\text{MC}$, for different values of window size, $l$, and fixed value of $k$, $k=8$. The results were averaged over 100 independent runs.}
\label{fig:k8windows}
\end{minipage}
\end{figure}

\Fref{fig:k8windows} demonstrates how the normalised entropy depends on the window size for a particular value of $k$. The information entropy is smaller for the final well-organised state in compare with the initial jammed state. More significant decrease of the information entropy was observed for the window size of order of $k$. Non-monotonic behaviour of the entropy for large window size ($l=32$) is noteworthy and not quite clear.

\begin{figure}
\centering
\includegraphics[width=\linewidth]{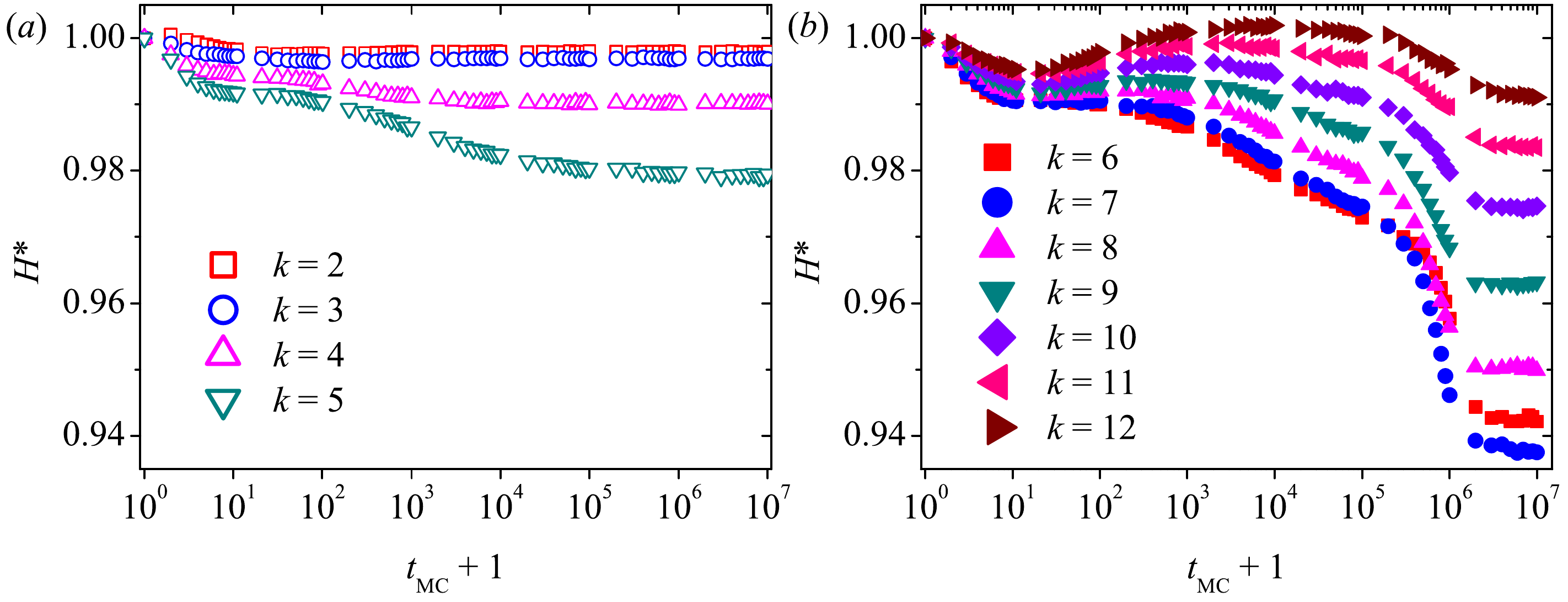}
\caption{Examples of the normalised entropy, $H^\ast$  vs MC steps, $t_\text{MC}$, for different values of  $k$, and fixed value of window size $l$, $l=8$. The results were averaged over 100 independent runs. ($a$)~no stripe domains were observed, ($b$) stripe domains were observed.}
\label{fig:entropyallk}
\end{figure}

\Fref{fig:entropyallk} demonstrates how the normalised entropy varies with time for different values of $k$. \Fref{fig:entropyallk}($a$) corresponds to the values of $k$ when no stripe domains are observed in the final state whilst \fref{fig:entropyallk}($b$) corresponds to the values of $k$ when stripe domains formed. Different behaviour of curves is clearly visible.

\section{Conclusion}\label{sec:concl}
Random walk of rectangular rigid particles $1 \times k$ on square lattice with unit sized cells and  PBCs may produce macroscopic patterns~\cite{Lebovka2017PRE,Tarasevich2017JSM,Patra2018PRE}. Due to hard-core interaction between the particles~\eref{eq:HCI}, the internal energy associated with particle positions is absent, hence, the free energy depends only on entropy. Any structural changes of the system should be treated as entropy-driven, i.e., an increase in macroscopic order is driven by an increase of microscopic disorder~\cite{Frenkel1993}. We found that random walk of rodlike particles turns initial nonequilibrium jammed state produced by RSA into a state with more movable particles. This transition is characterised both by increase in the DOF and by decrease in the information entropy. When $k > 5$, the pattern formation can be observed, viz., the final state of the system is the diagonal stripe domains. The information entropy is calculated from the frequency of occurrence of representatives of equivalence classes of two-dimensional words over a three-character alphabet in a window of fixed length. At the stage of pattern formation, decrease in the information entropy is accompanied by the increase in the number of  the DOF. We propose the following explanation of this effect. The increase in the DOF means that most $k$-mers (value of DOF is $\approx 0.75$) have at least one free cell connected to their end. This leads to the fact that, when a window shifts by one position, the number of free cells observed in the window varies little~---they are ``tied'' to $k$-mers. Consequently, the variety of observed representatives of the equivalence classes of two-dimensional words falls, which leads to a decrease in the information entropy. This, in part, also explains the fact that two-dimensional windows of size $l \gtrapprox k$ are most sensitive to changes of the microstructure of the system.

\ack The reported study was funded by RFBR according to the research project No.~18-07-00343.

\section*{References}
\bibliography{Entropy}

\providecommand{\newblock}{}
\begin{thebibliography}{10}
\expandafter\ifx\csname url\endcsname\relax
  \def\url#1{{\tt #1}}\fi
\expandafter\ifx\csname urlprefix\endcsname\relax\def\urlprefix{URL }\fi
\providecommand{\eprint}[2][]{\url{#2}}

\bibitem{Aranson2006RMP}
Aranson I~S and Tsimring L~S 2006 {\em Rev. Mod. Phys.\/} {\bf 78}(2) 641--692
  ISSN 0034-6861

\bibitem{Boerzsoenyi2013SM}
B{\"o}rzs{\"o}nyi T and Stannarius R 2013 {\em Soft Matter\/} {\bf 9}
  7401--7418 ISSN 1744-6848

\bibitem{Muller2015PRE}
M\"uller T, de~las Heras D, Rehberg I and Huang K 2015 {\em Phys. Rev. E\/}
  {\bf 91}(6) 062207 ISSN 2470-0045

\bibitem{Gonzalez2017SM}
Gonz\'{a}lez-Pinto M, Borondo F, Martinez-Rat\'{o}n Y and Velasco E 2017 {\em
  Soft Matter\/} {\bf 13}(14) 2571--2582 ISSN 1744-6848

\bibitem{Lebovka2017PRE}
Lebovka N~I, Tarasevich Y~Y, Gigiberiya V~A and Vygornitskii N~V 2017 {\em
  Phys. Rev. E\/} {\bf 95}(5) 052130 ISSN 2470-0045

\bibitem{Tarasevich2017JSM}
Tarasevich Y~Y, Laptev V~V, Burmistrov A~S and Lebovka N~I 2017 {\em J. Stat.
  Mech.\/} {\bf 2017} 093203 ISSN 1742-5468

\bibitem{Tarasevich2018JPhCSpacking}
Tarasevich Y~Y, Laptev V~V, Chirkova V~V and Lebovka N~I 2018 {\em J. Phys.
  Conf. Ser.\/} ISSN 1742-6588 accepted, arXiv:1803.09409

\bibitem{Evans1993RMP}
Evans J~W 1993 {\em Rev. Mod. Phys.\/} {\bf 65}(4) 1281--1329 ISSN 0034-6861

\bibitem{Frenkel1993}
Frenkel D 1993 {\em Physics World\/} {\bf 6} 24 ISSN 0953-8585

\bibitem{Mitescu1983}
Mitescu C~D and Roussenq J 1983 Diffusion on percolation clusters {\em
  Percolation Processes and Structures\/} ({\em Ann. Israel Phys. Soc.\/}
  vol~5) ed Deutscher G, Zallen R and Adler J (Bristol: Adam Hilger) pp 81--100
  ISBN 978-0852744772

\bibitem{Selinger1990PRA}
Selinger R~B and Stanley H~E 1990 {\em Phys. Rev. A\/} {\bf 42}(8) 4845--4852
  ISSN 1050-2947

\bibitem{Patra2018PRE}
Patra S, Das D, Rajesh R and Mitra M~K 2018 {\em Phys. Rev. E\/} {\bf 97}(2)
  022108 ISSN 2470-0045

\bibitem{LandauBinder}
Landau D~P and Binder K 2014 {\em A Guide to Monte Carlo Simulations in
  Statistical Physics\/} 4th ed (Cambridge: Cambridge University Press) ISBN
  9781107074026

\bibitem{Linares2008JSM}
Linares D~H, Rom\'{a} F and Ramirez-Pastor A~J 2008 {\em J. Stat. Mech. Theory
  E.\/} {\bf 2008} P03013 ISSN 1742-5468

\bibitem{Vogel2017PRE}
Vogel E~E, Saravia G and Ramirez-Pastor A~J 2017 {\em Phys. Rev. E\/} {\bf
  96}(6) 062133 ISSN 2470-0045

\bibitem{Smetanin2016MBB}
Smetanin Y~G, Ulyanov M~V and Pestova A~S 2016 {\em Math. Biol. Bioinf.\/} {\bf
  11} 114--126 ISSN 1994-6538 in Russian

\end{thebibliography}

\end{document}